\documentclass[oneside]{article}
\usepackage[T1]{fontenc}
\usepackage{authblk}
\usepackage{float}
\usepackage{amsmath}
\usepackage{graphicx}
\usepackage{xfrac}
\usepackage[english]{babel}
\usepackage{lmodern}
\usepackage[T1]{fontenc}
\usepackage[latin9]{inputenc}
\usepackage{mathtools}
\usepackage{graphicx}
\usepackage{esint}
\usepackage[unicode=true,
 bookmarks=false,
breaklinks=false,pdfborder={0 0 1},backref=false,colorlinks=false]
 {hyperref}
\usepackage[a4paper, total={210mm,297mm},margin=2.0cm]{geometry}

\usepackage{datetime}
\newdate{date}{30}{10}{2017}

\title{\textbf{Regularized lattice Boltzmann Multicomponent models for low Capillary and Reynolds microfluidics flows.}}

\author[1]{Andrea Montessori \thanks{Electronic address: \texttt{and.montessori@gmail.com}; Corresponding author}}

\author[2]{Marco Lauricella}

\author[1]{Michele La Rocca}

\author[2,3]{Sauro Succi}

\author[4]{Elad Stolovicki}

\author[4]{Roy Ziblat}

\author[5,4] {David Weitz}

\affil[1]{Department of Engineering, University of Rome \textquotedbl{}Roma Tre,\textquotedbl{} Via della Vasca Navale 79, 00141 Rome, Italy}

\affil[2]{Istituto per le Applicazioni del Calcolo CNR, Via dei Taurini 19,
00185 Rome, Italy}

\affil[3]{Institute for Applied Computational Science, Harvard John A. Paulson School of Engineering And Applied Sciences, Cambridge, MA 02138, United States}

\affil[4]{School of Engineering and Applied Sciences, Harvard University, McKay 517 Cambridge, MA 02138, USA}

\affil[5]{Department of Physics, and School of Engineering and Applied Sciences, Harvard University, Pierce 231, 29 Oxford St. Cambridge, MA 02138, USA}

\date{\displaydate{date}}

\begin{document}

\maketitle

\begin{abstract}

We present a regularized version of the color gradient lattice Boltzmann (LB) scheme 
for the simulation of droplet formation in microfluidic devices of experimental relevance.   
The regularized version is shown to provide computationally efficient access  to Capillary 
number regimes relevant to droplet generation via microfluidic devices, such as flow-focusers
and the more recent microfluidic step emulsifier devices.

\end{abstract}

\section{Introduction}

In the last two decades, microfluidic devices have gained a prominent role in several fields 
of research, from basic fluid dynamics to material science, biomedicine, as well as industrial applications \cite{anna2016droplets,vladisavljevic2013industrial,theberge2010microdroplets}.
In the early 2000s, several pioneering works showed the potential of such devices for generating 
droplets at the microscale with unprecedented degree of uniformity and rational design, thereby
establishing the basis of the lab-on-a-chip concept \cite{utada2005monodisperse,anna2003formation,ganan2001perfectly,thorsen2001dynamic}.
Nowadays, many publications show that the microfluidics has surged well beyond the proof-of-concept paradigm,
proving the viability of the new approach through substantial contributions to chemistry, biology, medicine,  3d-printing, to name but a few \cite{sackmann2014present,rodriguez2015generation,elvira2013past,mazutis2013single,bhattacharjee2016upcoming}.
Due to their ease of fabrication via soft lithography methods \cite{mcdonald2000fabrication,whitesides2001flexible}, microfluidic devices 
are intensely exploited for the study and manipulation of fluids at the submillimeter length scale.
In particular, microfluidic devices have been successfully employed for producing porous 
scaffolding materials with an accurate control over scaffold specifications, such as pore size, shape, 
distribution and interconnectivity \cite{costantini2014highly,oh2015injectable}.

In such context, droplet generation units are the main component
to produce emulsion templating porous materials by means of microfluidic devices.
Several droplet-based microfluidic chips include at least one droplet generation unit
within different geometries, alongside with droplet splitting/merging units 
(e.g., flow focusing, coflow,  T-, X-, and Y-junctions).
Experiments have driven many of the advances in the field.
Nonetheless, many quantities of design interest lie still beyond experimental reach, thereby 
precluding a complete understanding of the basic physics of droplet generation by experimental 
means and thus holding back  further progress in the operation and optimization of microfluidic devices.

Models and simulations may provide valuable insights into basic microfluidic
mechanisms. More specifically, computational studies can help elucidating the nature of optimal flow conditions 
in terms of geometrical and physico-chemical properties, thus facilitating a rational design of the final product.

Over a decade ago, different numerical methods focused on the breakup mechanisms \cite{dupin2006simulation,de2006modeling},
characterizing droplet formation in terms of the relevant dimensionless parameters \cite{de2008transition,liu2011droplet}.
In particular, it was noted that by varying volume flow rates of the dispersed and continuous phases, and therefore
changing the Reynolds and Capillary numbers, three distinct regimes of 
formation of droplets can be identified: {\it squeezing, dripping and jetting}.
All three regimes are consistent with experimental observations \cite{utada2005monodisperse,garstecki2006formation,utada2007dripping}.
The lattice Boltzmann (LB) method has played a major role in the simulation of droplet formation across a wide variety of 
microfluidic cross-junctions \cite{liu2011droplet,wang2011lbm,falcucci2011modern,yan2012numerical,gupta2016lattice}.

The LB method  is known to experience stability and efficiency limitations at both low and high 
viscosities \cite{aidun2010lattice}; low viscosities are known to threaten numerical stability, while
large ones undermine the very hydrodynami limit of the LB scheme, due to the onset of strong non-equilibrium effects.
 The latter are raise a specific concern for the simulation microfluidic configurations.

Different strategies can be employed to mitigate the above constraints:  the multirelaxation-time (MRT) method \cite{higuera1989lattice},
and a regularized version (REG) of the standard single-relaxation-time (SRT) LB scheme \cite{zhang2006efficient,latt2006lattice},
also known as Regularized lattice Bhatnagar-Gross-Krook model, as well as the entropic version of the LB method \cite{ansumali2006entropic}.

In this paper, we investigate and demonstrate the benefits of the regularization procedures, as
applied to the color gradient model introduced by Leclaire and co-workers \cite{leclaire2011isotropic,leclaire2012numerical}, 
for the simulation of microfluidic devices.

Indeed, the REG approach filters out the non-hydrodynamic modes, also known as ghost-modes,
originating from non-equilibrium effects stemming from free molecular motion  between two subsequent collisions 
\cite{latt2006lattice,montessori2014regularized,montessori2015lattice,falcucci2017effects}.

As detailed in the sequel, this proves particularly useful for microfluidic applications characterised by 
low Capillary numbers.

The paper is organized as follows. In Sec. II the lattice Boltzmann equation with the BGK collisional operator is described, together 
with the colour gradient model for simulating multicomponent fluids and the regularization algorithm. 
In Sec. III the regularization algorithm is commented and its benefits for LB simulation in microfluidics context are highlighted. 
Section IV presents the results of flow-focusing simulations in two spatial dimensions as well as 
preliminary three-dimensional simulations of the newly proposed "volcano" micro devices.
Finally, a summary is provided in Sec. V.

\section{Methods}

The LB immiscible multi-component model is based on the following lattice Bhatnagar-Gross-Krook (BGK) equation:
\begin{equation}
f_{i}^{k} \left(\vec{x}+\vec{c}_{i}\Delta t,\,t+\Delta t\right) =f_{i}^{k}\left(\vec{x},\,t\right)+\Omega_{i}^{k}( f_{i}^{k}\left(\vec{x},\,t\right)),
\end{equation}
where $f_{i}^{k}$ is the discrete distribution function, representing
the probability of finding a particle of the $k-th$ component at position $\vec{x}$ and time
$t$ with discrete velocity $\vec{c}_{i}$ . 
The lattice time step is taken equal to 1, and $i$ the index spans the lattice discrete directions $i = 0,...,b$ \cite{leclaire2012numerical}.
The density $\rho^{k}$ of the $k-th$ fluid component is given by the zeroth moment of the distribution functions
\begin{equation}
\rho^{k}\left(\vec{x},\,t\right) = \sum_i f_{i}^{k}\left(\vec{x},\,t\right),
\end{equation}
while the total momentum $\rho \vec{u} $ is defined by the first order moment:
\begin{equation}
\rho \vec{u} = \sum_i  \sum_k f_{i}^{k}\left(\vec{x},\,t\right) \vec{c}_{i}.
\end{equation}
The collision operator $\Omega_{i}^{k}$ results from the combination of three sub-operators, namely \cite{Krafczyk2002,leclaire2012numerical}
\begin{equation}
\Omega_{i}^{k} = \left(\Omega_{i}^{k}\right)^{(3)}\left[\left(\Omega_{i}^{k}\right)^{(1)}+\left(\Omega_{i}^{k}\right)^{(2)}\right].
\end{equation}
Here, $\left(\Omega_{i}^{k}\right)^{(1)}$ is the standard BGK operator for the $k-th$
component, accounting for relaxation towards a local equilibrium
\begin{equation}
\left(\Omega_{i}^{k}\right)^{(1)} f_{i}^{k}\left(\vec{x},\,t\right) = f_{i}^{k}\left(\vec{x},\,t\right) - \omega_{k}\left(f_{i}^{k}\left(\vec{x},\,t\right)-f_{i}^{k,eq}\left(\vec{x},\,t\right)\right),
\end{equation}
where $\omega_{k}$ is the relaxation rate, and $f_{i}^{k,eq}\left(\vec{x},\,t\right)$ denotes local equilibria, defined by
\begin{equation}
f_{i}^{k,eq}\left(\vec{x},\,t\right)=  \rho^{k} \left[ \phi_{i}^{k} + w_i \left(  \frac{\vec{c}_i \cdot \vec{u}}{c_s^2}  + \frac{(\vec{c}_i \cdot \vec{u} )^2}{2c_s^4} -  \frac{( \vec{u} )^2}{2c_s^2} \right) \right].
\end{equation}
Here, $w_i$ are weights of the discrete equilibrium distribution functions (e.g. we use a standard D2Q9 lattice), 
$c_s$ is the lattice sound speed, and $\phi_{i}^{k}$ takes values in D2Q9 lattice
\begin{equation}
\phi_{i}^{k}=\begin{cases}
\alpha_{k}, & i=1,\\
\left(1-\alpha_{k}\right)/5, & i=2,4,6,8,\\
\left(1-\alpha_{k}\right)/20, & i=3,5,7,9,
\end{cases}
\end{equation}
where we number $ i=2,4,6,8$ the nearest-neighbour lattice displacements, and $ i=3,5,7,9$ the diagonal ones.
In the above expression, $\alpha_{k}$ is a free parameter, modulating the density ratio $\gamma_k$ of
the $k-th$ component with respect to the others \cite{grunau1993lattice}, as well as tuning its relative pressure
\begin{equation}
p^k=\frac{3\rho^{k}\left(1-\alpha_{k}\right) }{5}.
\end{equation}

In this model, $\left(\Omega_{i}^{k}\right)^{(2)}$ is a perturbation operator, modelling the surface tension of the $k-th$ component.
Denoting by $\vec{F}$ the colour gradient in terms of the colour difference, this term reads
\begin{equation}
\left(\Omega_{i}^{k}\right)^{(2)} f_{i}^{k}\left(\vec{x},\,t\right)= f_{i}^{k}\left(\vec{x},\,t\right)+\frac{A_k}{2}|\vec{F}| \left[w_i \frac{(\vec{F} \cdot \vec{c}_i)^2}{|\vec{F}|^2} -B_i \right],
\end{equation}
with the free parameters $A_k$ modelling the surface tension, and $B_k$ a parameter depending on the
chosen lattice \cite{reis2007lattice,leclaire2011isotropic}.

The above operator models the surface tension, but it does not guarantee the immiscibility between the various components. 
In order to minimize the mixing of the fluids, a recolouring operator $\left(\Omega_{i}^{k}\right)^{(3)}$  is introduced.
Following the approach in Ref. \cite{leclaire2011isotropic}, denoted by $\zeta$ and $\xi$ two immiscible fluids, the recolouring operators 
for the two fluids read as follows"
\begin{equation}
\begin{aligned}
\left(\Omega_{i}^{\zeta}\right)^{(3)} &= \frac{\rho^{\zeta}}{\rho}f_i + \beta \frac{\rho^{\zeta} \rho^{\xi}}{\rho^2} \cos(\phi_i) \sum_k f_i^{k,eq} (\rho^k , 0, \alpha_k) \\
\left(\Omega_{i}^{\xi}\right)^{(3)} &= \frac{\rho^{\xi}}{\rho}f_i - \beta \frac{\rho^{\zeta} \rho^{\xi}}{\rho^2} \cos(\phi_i) \sum_k f_i^{k,eq} (\rho^k , 0, \alpha_k)
\end{aligned}
\end{equation}
where $\beta$ is a free parameter and $\cos(\phi_i)$ is the cosine of the angle between the colour gradient $\vec{F}$
and the lattice direction $\vec{c}_i$.
Note that $f_i^{k,eq} (\rho^k , 0, \alpha_k)$ stands for the equilibrium distributions of $k-th$ fluid
evaluated using the respective value of $\rho^k$, zero velocity, and $\alpha_k$.
In the above Eq, $f_i=\sum_k f_i^k$.
\begin{figure}
\centering
\includegraphics[width=\hsize]{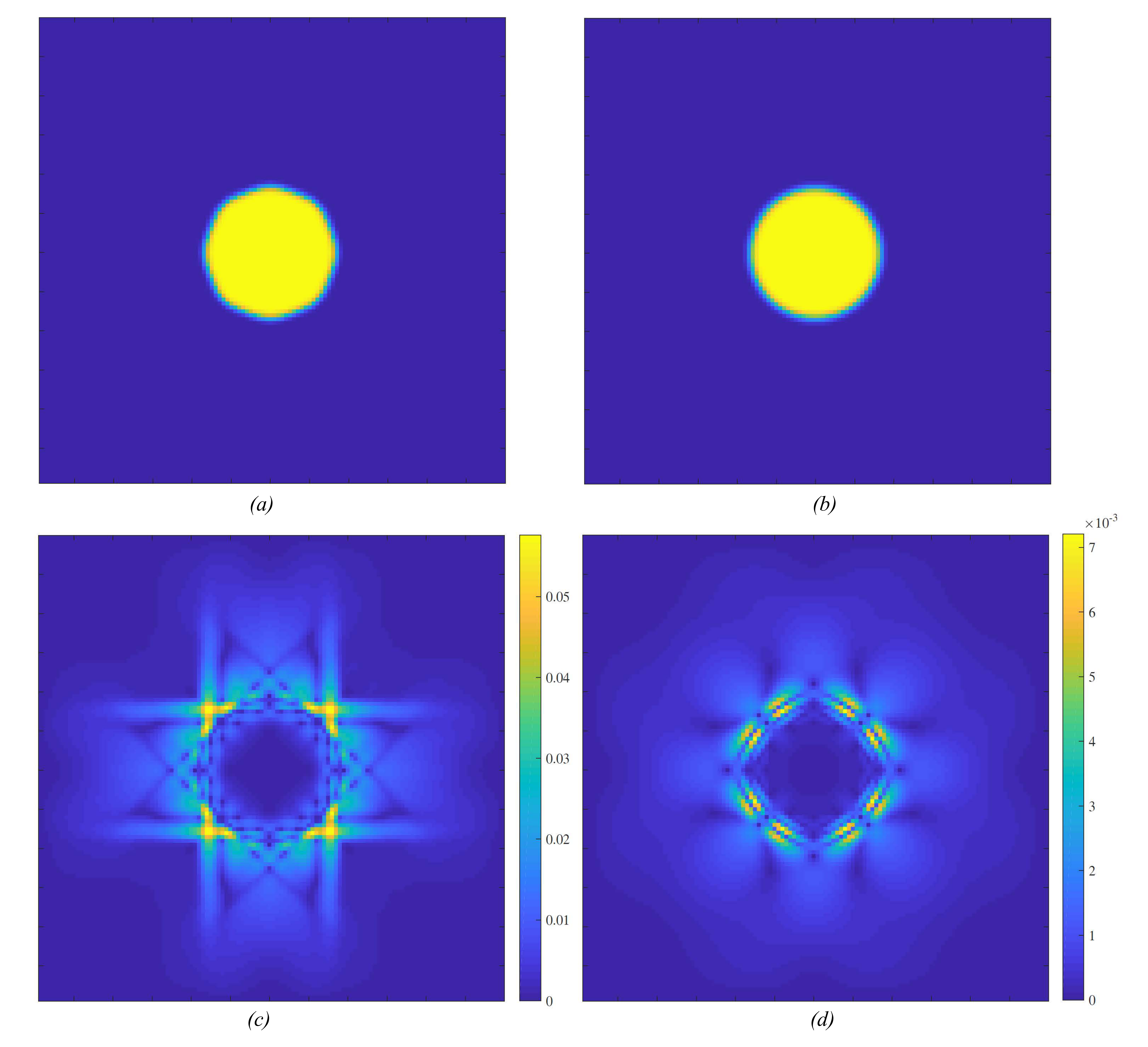}
\caption{\label{spurious} (a) and (b); density field of a resting droplet  immersed in a second component : (a) LBGK (b) Regularized LBGK. 
The relaxation times of the blue and yellow component were set, respectively,  to 1 and 5 ($\nu_B \sim 0.167$, $\nu_Y\sim 1.5$).
(b) and (c) velocity field (spurious currents) around the resting droplets for the non regularized (b) and Regularized (c) case.
It is evident that, for high viscosity values of the dispersed phase, a reduction of isotropy emerges that reflects into a non spherical shape of the resting droplet. 
An inspection of the flow field highlights that the droplet anisotropy is basically driven by a non-physical flow field around the droplet which is, in turn, caused by the presence of the ghost modes which are excited in the  under-relaxed regime ($\tau>1$). 
As one can see, the Regularization cures the loss of isotropy in under-relaxed LBGK, by suppressing 
the non-physical modes, as evidenced by the  circular shape of the droplet at rest and by the spurious flow field around 
the droplet, which is now isotropic and  roughly an order of magnitude smaller than in the plain LBGK case.}
\end{figure}
The LB colour gradient model has been enriched with the so called Regularization procedure \cite{montessori2014regularized,zhang2006efficient,falcucci2017effects}, namely a discrete Hermite projection of the post-collisional set of distribution functions onto a proper set of Hermite basis.
The main idea is to introduce a set of pre-collision distribution functions which are defined only in terms of the 
macroscopic hydrodynamic moments.
\begin{equation}
f_i^k(x_i+c_i \Delta t,t+\Delta t) = \mathcal{R} f_i^{k,neq}(x,t) \equiv h_i^{k,eq} -  \frac{\Delta t}{\tau} h_i^{k,neq}
\end{equation}
where $h_i^k$ is the hydrodynamic component of the full distribution $f_i^k$ (see \cite{zhang2006efficient}) for the $k-th$ colour, and $\mathcal{R}$ 
is the regularization operator. The above equation shows that the post-collision distribution, of a $4^{th}$-order isotropic lattice, 
is defined only in terms of the conserved and the transport hydrodynamic modes, 
namely density $\rho$, current $\rho \vec{u}$ and momentum-flux tensor $\mathbf{\Pi}=\sum_i f_i \vec{c_i}\vec{c_i}$.
The motivation for using the regularization procedure in the colour gradient model is detailed in the next Section.

\section{Regularized LB multicomponent approach for low Capillary and Reynolds microfluidics}

The Reynolds and the Capillary numbers are defined as follows:
\begin{equation}
\begin{aligned}
Ca &= \frac{\rho \nu U}{\sigma} \\
Re &= \frac{L U}{\nu}
\end{aligned}
\end{equation}
where $\nu$ is the kinematic viscosity of the fluid, $\sigma$ is the surface tension and $L$ and $U$ are respectively, the characteristic length and velocity.
Microfluidic flows in T-junctions and flow focusing devices are most often characterized by  $Re\sim1$ and $Ca<<1$.
For a typical droplet of diameter $D=10^{-4}$ (m), moving at a speed $U = 0.01$ (m/s) the Reynolds and Capillary
numbers are $Re=10$ and $Ca=10^{-4}$, where we have taken the density of water and a surface tension of $\sim 10mN/m$.  
The diffuse nature of the fluid-fluid interface in the multicomponent model employed in this work, poses some 
constraint on the ratio between the characteristic length scale of the problem, namely the droplet diameter $D$,  
and the width of the diffuse interface, $\delta$, known as Cahn number  $Cn=\delta/D$ \cite{cahn1958free}.
Since the interface width lies on nanometric scales, the Cahn number is usually very small, of the order of $10^{-4}$ or less.
Such scale separation is computationally unpractical, and LB simulations must typically operate at much higher
Cahn numbers, between $0.01 \div 0.1$, implying that the associated inaccuracies must be properly inspected.  
Given that the diffuse interface is about $5$ lattice units, by taking $D_{lb}=100$ (subscript lb denotes lattice units), 
we obtain $Cn=0.05$.
Realizing $Re=1$ with $Cn\sim0.05$ runs against numerical limitations of the LB method; for instance,
by choosing $U_{lb}=0.001$ and $\nu_{lb}=0.1$, one faces two inconveniences: first, very long simulation time
due to the small velocities, ($U_{lb}=0.001$ means thousands lattice time steps to cover one lattice spacing...),
second, a low signal/noise ratio due to spurious currents.
The alternative is to raise both the velocity and the viscosity, say $U_{lb}=0.01$ and $\nu_{lb}=1$; 
however, this implies large values of the relaxation time $\tau$, triggering correspondingly 
large non-equilibrium effects and ghost currents.
This is where the benefits of the Regularization technique take stage: by filtering out the ghost 
currents, the Regularized LB can operate at higher values of the droplet speed without incurring
into spurious currents and anomalous ghost effects, \cite{montessori2015lattice}.
\begin{figure*}
\centering
\includegraphics[scale=0.2]{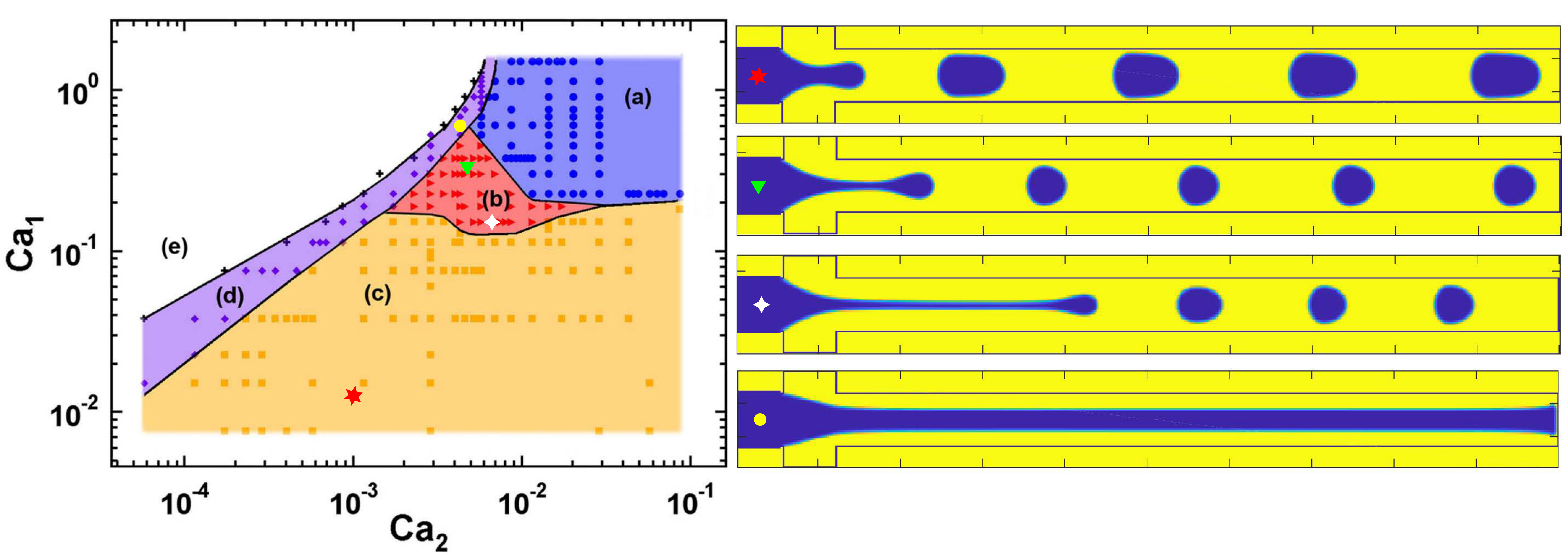}
\caption{\label{regimes}  Capillary number-based flow map with flow regimes reproduced by the regularized color gradient model (from the upper left to the lower left panel): Dripping,  jetting (second and third) and tubing.
The regularized model is capable of accurately predicting the different flow regimes in a microfluidic flow focusing device. The viscosities of the two fluids are $\nu_Y=0.167$(continuous phase) and $\nu_B \sim 0.5$ (dispersed phase), thus matching the viscosity ratio of the liquids employed in the experiments reported in \cite{cubaud2008capillary}.}
\end{figure*}
\section{Results}

In the following, we present two preliminary applications of the proposed scheme, namely the simulation
of droplet formation in standard flow-focussing microfluidic devices and to the recently proposed "volcano"
devices \cite{stolovicki2017throughput}.

\subsection{Flow-focussing devices}
We performed simulations of resting droplets of component one (coloured as yellow in Fig. \ref{spurious}) immersed
in a second component (blue in Fig. \ref{spurious} ) with same densities and different viscosities.
\begin{figure}
\centering
\includegraphics[scale=0.25]{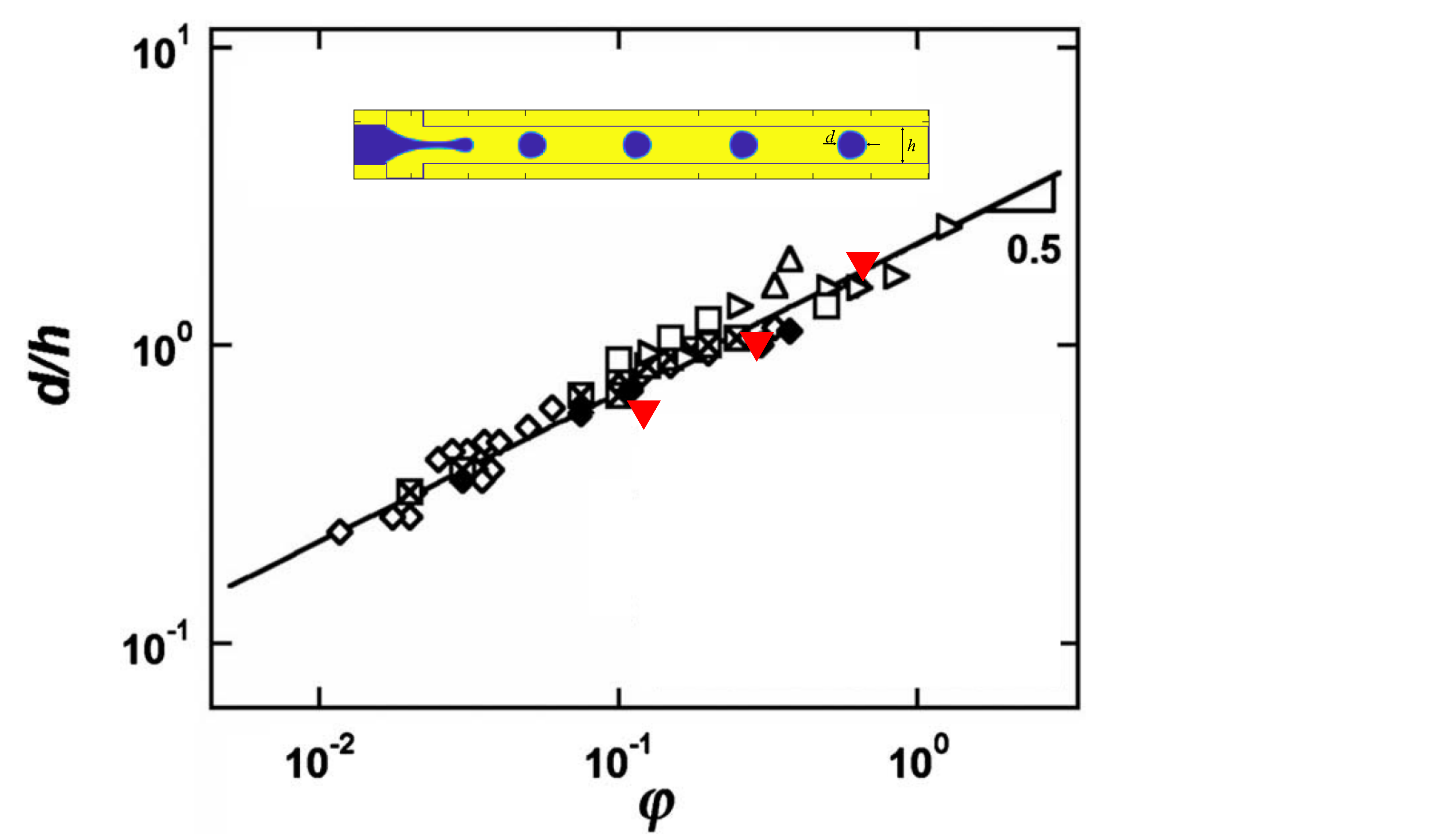}
\caption{\label{dovh} Normalized droplet diameter $d/h$ vs the flow rate ratio $\phi=Q_B/Q_Y$. Numerical results (Red triangles) are superimposed to the experimental curve of Cubaud et al. \cite{cubaud2008capillary}.
The diameters collapse on a single master curve, which scales with the flow rate like $Q_B/(2 Q_Y)^{0.5}$, as reported in \cite{cubaud2008capillary}.}
\end{figure}
The relaxation times of the blue and yellow components were set to $1$ and $5$ ($\nu_B \sim 0.167$, $\nu_Y\sim 1.5$), respectively.
It is worth to highlight that, for high viscosity values of the dispersed phase, sizeable non-isotropic effects arise, 
resulting in a non-spherical shape of the rest droplet. 
A more detailed inspection of the flow field shows that the droplet anisotropy is basically driven by a non-physical flow 
field around the droplet, caused by the presence of the ghost modes. 
The ghost modes are excited whenever under-relaxed ($\tau>1$) LBGK models are employed \cite{montessori2015lattice}. 
The regularization cures the loss of isotropy in under-relaxed LBGK, by suppressing the non-physical modes as evidenced by the  circular shape of the droplet at rest and by the spurious flow field around the droplet (see Fig. \ref{spurious} left panels) which is isotropic. Further, the maximum value of spurious currents is an order of magnitude smaller than in the plain LBGK case.
We then run simulations of a flow focusing device and compared our results with experimental data available in literature. 
The viscosities of the two fluids are $\nu_Y=0.167$(continuous phase) and $\nu_B \sim 0.5$ (dispersed phase), thus matching the viscosity ratio of the liquids employed in the experiments of Ref. \cite{cubaud2008capillary}.
Even in this case, the relaxation time of the dispersed phase (blue in Fig. \ref{regimes}) is greater than one, providing an out
of equilibrium regime as in the previous case.
\begin{figure}
\centering
\includegraphics[scale=0.26]{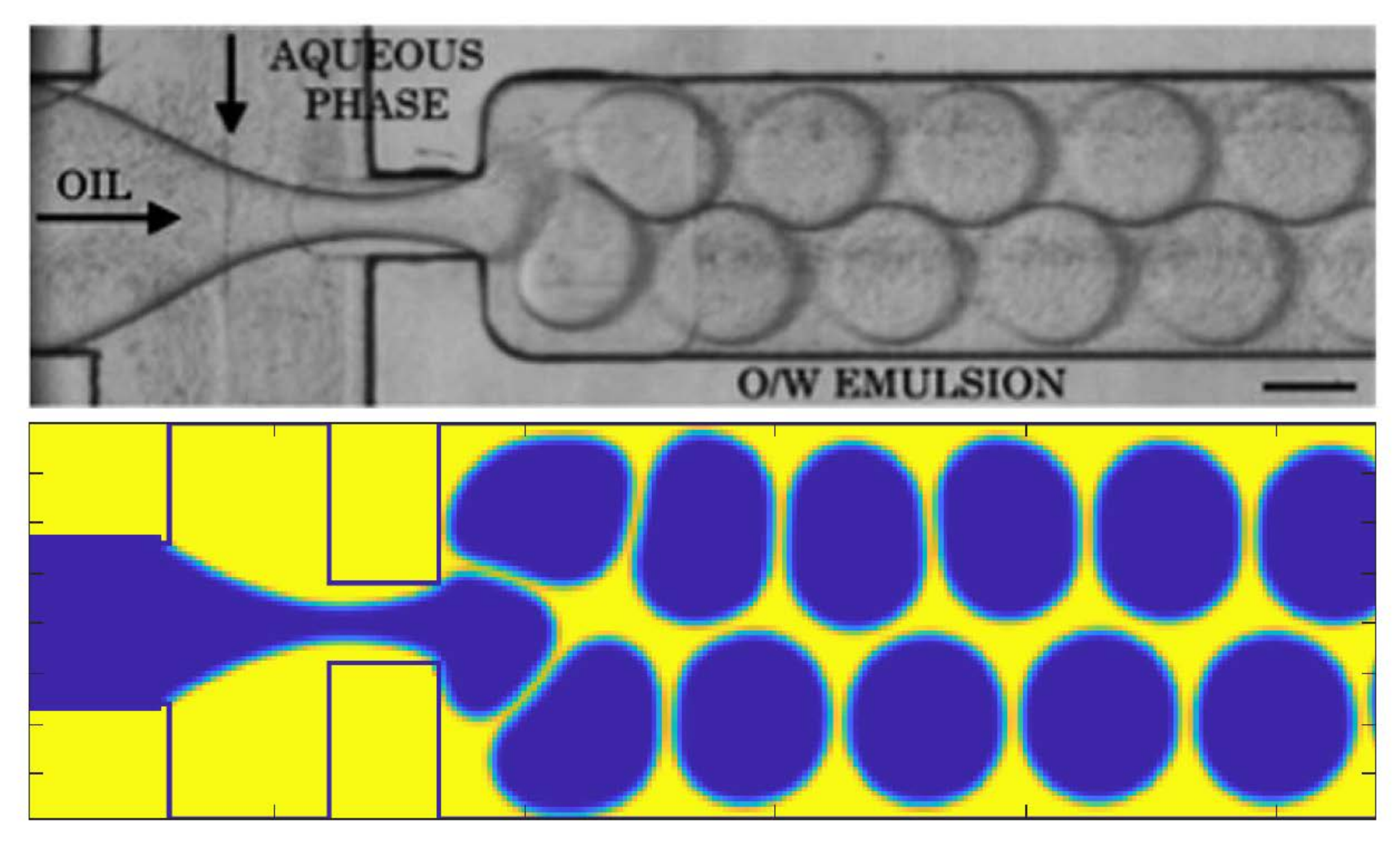}
\caption{\label{owemulsion}  Prospective application of regularized colour gradient model augmented with arrested coalescence algorithm.
The model allows for stable simulation of mono-disperse droplets. This opens the way to the simulation of mono-dispersed oil-water emulsions. Future applications of this model will allow to  identify optimal operational regimes, capable of delivering droplet configurations of high regularity, both in size and connectivity. Upper panel containing the experimental data is reported from Ref. \cite{costantini2014highly} }
\end{figure}
In Fig. \ref{regimes} we report the Capillary number-based flow map with flow regimes observed by Cubaud et al. \cite{cubaud2008capillary}.
The regularized colour gradient is clearly able to reproduce the different flow regimes in such
microfluidic flow-focusing device, correctly predicting dripping, jetting and tubing flow configurations
at different Capillary numbers.
Next, we computed the normalized droplet diameter $d/h$, being $h$ the height of the microfluidic channel, versus the flow 
rate ratio ($Q_B/Q_Y$) and compare with experimental data.
The numerical results, in agreement with Ref. \cite{cubaud2008capillary}, collapse on a single master curve, which
scales with the flow rate like $Q_B/(2 Q_Y)^{0.5}$ (see Fig. \ref{dovh}).
As a prospective application, the regularized approach was used to simulate an oil/water emulsion in a flow-focusing device.
In order to obtain a mono-dispersed emulsion, the regularized colour gradient approach has been augmented with
an algorithm aimed at suppressing coalescence between the droplets of the dispersed phase (oil).
The results show that the model allows for stable simulation of mono-disperse droplets, well reproducing the experimental data (see upper panel of Fig. \ref{owemulsion}). This opens the way to the simulation of mono-dispersed oil-water emulsions. 
Future applications of this model will allow to  identify optimal operational regimes capable of delivering droplet configurations of high regularity, both in size and connectivity. 

\subsection{Microfluidic volcanos}

As a further application, we report some preliminary simulations 
of a new class of step emuldification devices, called volcano \cite{stolovicki2017throughput}, which are based on the idea
of preventing the obstruction of the nozzles from the droplets via buoyancy effects. 
These devices are expected to  enhance the yield of highly mono-dispersed water/oil emulsion, 
which is highly desirable for most industrial purposes.\\
The volcano device made from polydimethylsiloxane is used here for producing water in oil emulsions. 
\begin{figure*}
\centering
\includegraphics[scale=0.07]{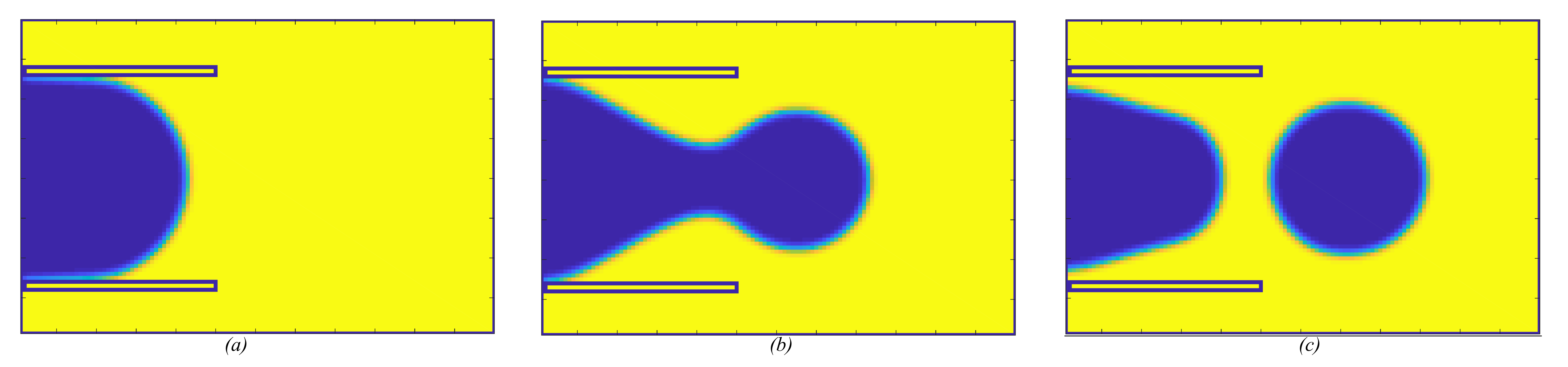}
\caption{\label{sequencevolc}  Density field at the $x-z$ Mid-plane of the nozzle of the volcano device. 
The width of the channel $w=700 \mu m $ and the  height $h= 70 \mu m$ ($h/w=1/10$), corresponding to 
Capillary number is $Ca=1.4\times 10^{-2}$ 
After the break up, the droplet diameter is $D=350 \mu m$, so that $D/w=0.5$, in good agreement with the 
experimental findings on the volcano device ($w/h=8$, $D/w=0.58$, $Ca=10^{-2}$)) \cite{stolovicki2017throughput}.
}
\end{figure*}
The water flows through the device inlet, and splits into hundreds of step-emulsifier nozzles with rectangular cross section.  
The device is submerged in a quiescent oil reservoir, each nozzles producing
a stream of micron-sized droplets\\
As a preliminary step, we simulated a single-nozzle device. 
Upon matching the governing dimensionless groups (Capillary and Weber number) and the characteristic geometrical 
ratio $h/w$ (see the caption of fig. \ref{sequencevolc} for the values of the physical paramters), we are able 
to simulate the droplet break up.
After the break up, the droplet diameter is $D\sim 350\mu m$, corresponding to  $D/w \sim 0.5$, in good agreement with the 
experimental findings on the volcano device (\cite{stolovicki2017throughput}) ($w/h=8$, $D/w=0.58$, $Ca=10^{-2}$)).
A thorough investigation of the microfluidics of volcano devices is currently underway,  and will make 
the subject of future communications.
\section{Summary}
Summarising, we have presented a novel variant of the Lattice Boltzmann method for
multiphase flows, based on the regularisation of the colour-gradient scheme, augmented
with a color-swap algorithm to mimic the effect of intermolecular repulsion, so as to tame droplet coalescence. 
The new scheme has been applied to the simulation of droplet production in flow-focussing
micro-devices, finding satisfactory agreement with the existing literature, both in terms
of predicting the transition from dripping-jetting-tubing regimes, and also in terms of
spacetime patterns of the droplet configurations in experimental devices.
Moreover, we have also presented preliminary three-dimensional simulations of an
alternative step emulsifier device, known as  volcano device, which is based on the idea of 
promoting nozzles cleaning via buoyancy effects.
If successfully demonstrated, volcano devices may offer substantial advantages in terms
of droplet rate production, degree of mono-dispersity and morphological regularity. 

\section*{Acknowledgments}

The research leading to these results has received
funding from the European Research Council under the European
Union's Horizon 2020 Framework Programme (No. FP/2014-
2020)/ERC Grant Agreement No. 739964  (\textquotedbl{}COPMAT\textquotedbl{}).

\end{document}